\documentstyle[11pt]{article}
\normalbaselines

\begin{document}
\bibliographystyle{unsrt}

\newtheorem{theorem}{Theorem}
\newtheorem{lemma}{Lemma}
\newtheorem{proposition}{Proposition}
\newtheorem{corollary}{Corollary}

\def\bea*{\begin{eqnarray*}}
\def\eea*{\end{eqnarray*}}
\def\ba{\begin{array}}
\def\ea{\end{array}}
\count1=1
\def\be{\ifnum \count1=0 $$ \else \begin{equation}\fi}
\def\ee{\ifnum\count1=0 $$ \else \end{equation}\fi}
\def\ele(#1){\ifnum\count1=0 \eqno({\bf #1}) $$ \else \label{#1}\end{equation}\fi}
\def\req(#1){\ifnum\count1=0 {\bf #1}\else \ref{#1}\fi}
\def\bea(#1){\ifnum \count1=0   $$ \begin{array}{#1}
\else \begin{equation} \begin{array}{#1} \fi}
\def\eea{\ifnum \count1=0 \end{array} $$
\else  \end{array}\end{equation}\fi}
\def\elea(#1){\ifnum \count1=0 \end{array}\label{#1}\eqno({\bf #1}) $$
\else\end{array}\label{#1}\end{equation}\fi}
\def\cit(#1){
\ifnum\count1=0 {\bf #1} \cite{#1} \else 
\cite{#1}\fi}
\def\bibit(#1){\ifnum\count1=0 \bibitem{#1} [#1    ] \else \bibitem{#1}\fi}
\def\ds{\displaystyle}
\def\hb{\hfill\break}
\def\comment#1{\hb {***** {\em #1} *****}\hb }

\newcommand{\TZ}{\hbox{\bf T}}
\newcommand{\MZ}{\hbox{\bf M}}
\newcommand{\ZZ}{\hbox{\bf Z}}
\newcommand{\NZ}{\hbox{\bf N}}
\newcommand{\RZ}{\hbox{\bf R}}
\newcommand{\CZ}{\,\hbox{\bf C}}
\newcommand{\PZ}{\hbox{\bf P}}
\newcommand{\QZ}{\hbox{\bf Q}}
\newcommand{\HZ}{\hbox{\bf H}}
\newcommand{\EZ}{\hbox{\bf E}}
\newcommand{\GZ}{\,\hbox{\bf G}}

\font\germ=eufm10
\def\goth#1{\hbox{\germ #1}}
\vbox{\vspace{38mm}}

\begin{center}
{\LARGE \bf Algebraic Geometry  and Hofstadter Type
Model}\\[10 mm]
Shao-shiung Lin \\ {\it Department of Mathematics,\\ 
Taiwan University \\ Taipei, Taiwan \\ 
(e-mail: lin@math.ntu.edu.tw ) }\\[3 mm] Shi-shyr
Roan
\footnote{Supported in part by the NSC grant of Taiwan . }
\\{\it Institute of Mathematics \\ Academia Sinica \\ 
Taipei , Taiwan \\ (e-mail: maroan@ccvax.sinica.edu.tw)} 
\\[35mm]
\end{center}

\begin{abstract}
In this report, we study the algebraic geometry
aspect of Hofstadter type models through the 
algebraic Bethe equation. In the diagonalization
problem of certain Hofstadter type Hamiltonians,
the Bethe equation is constructed by using the
Baxter vectors on a high genus spectral curve. 
When the  spectral
variables lie on rational curves, we obtain the
complete and explicit solutions of the
polynomial Bethe equation; the relation with the
Bethe ansatz of polynomial roots is discussed. 
Certain 
algebraic geometry properties of Bethe
equation on  the high genus algebraic curves  are 
discussed  in cooperation  with the
consideration of the physical model.
\end{abstract}

\section{Introduction}
It is known for the past decade that algebraic
geometry has played a certain intriguing role in
certain 2-dimensional solvable statistical
lattice models, a notable example would be the
chiral Potts $N$-state integrable model (see
e.g., \cite{AMP} \cite{BBP} and references
therein). In the note, we report the algebraic
geometry aspect of another 
model of physical interest in 
solid state physics. In the early 90's, 
motivated by the work of Wiegmann and Zabrodin
\cite{WZ} on the appearance of 
$U_q(sl_2)$ symmetry in problems of magnetic
translation, Faddeev and Kashaev \cite{FK}
pursued the diagonalization problem on the
following Hamiltonian by the quantum
transfer matrix  method which was developed by
the Leningrad school in the early eighties:
\be
H_{FK}= \mu ( \alpha U + \alpha^{-1} U^{-1} ) + \nu (
\beta V + \beta^{-1} V^{-1}) + \rho ( \gamma W +
\gamma^{-1} W^{-1} ) \ ,
\ele(HFK)
where
$U, V, W$ are unitary operators with the
Weyl commutation relation for a primitive $N$-th
root of unity $\omega$ and the
$N$-th power identity property, $
UV= \omega VU$, $ VW=
\omega WV$, $ WU= \omega UW$; $ U^N= V^N = W^N
=1$. 
As a special limit case for $\rho =0$, the model
is reduced to the (rational flux) Hofstadter
Hamiltonian, a model possessing several physical
interpretations with the history which can  trace
back to the work of Peierls \cite{P} on Bloch
electrons in  metals
with the presence of a constant external magnetic field.
By the pioneering works
of the 50s and 60s \cite{A} \cite{C} \cite{Har}
\cite{L} \cite{Z}, the role of magnetic
translations was found, and it began a systematic
study of this 2D lattice model. In 1976,
Hofstadter \cite{H} found
  the butterfly figure of the
spectral band versus the magnetic flux which exhibits a
beautiful fractal picture. Here the phase of 
$\omega$ represents  the magnetic flux (per
plaquette). In \cite{FK}, a
general frame work to determine the
eigeinvalues of certain quantum chains appeared
in the transfer matrix was presented.  The
method relies on a special  monodromy
solution of the Yang-Baxter equation for the
six-vertex
$R$-matrix; this solution appeared also in the 
study of chiral Potts model \cite{BBP}. 
For a finite size
$L$, the trace of the monodromy matrix
gives rise to the transfer matrix acting on the
quantum space $\stackrel{L}{\otimes}\CZ^N$; while
the Hofstadter type Hamiltonian (\req(HFK))
can be realized in the case $L=3$. In general, the
diagonalization problem of the transfer matrix
can be formulated into the Bethe equation 
through the Baxter vector\footnote{It is also 
called as the "Baxter vacuum state" in other
literature. }, visualized on a
"spectral" curve  associated to the corresponding
model. In \cite{LR}, we presented a
detailed and rigorous mathematical study on 
the Bethe equation associated to the Hofstadter
type model. In particular, we obtained the
complete solution of the Bethe equation for
models with rational spectral curves for $L
\leq 3$, among which a special Hofstadter  type
of $H_{FK}$ in \cite{FK} is included, and further
expended to  all the other sectors. In this note,
we explain the main results we have obtained in
\cite{LR};
detailed derivations, as well as extended
references to the literature, may be found in
that work. 

This paper is organized as follows. 
In Sect. 2, 
we  first recall  results in transfer matrix
relevant to our discussion; then introduce
the Bethe
equation (or Baxter $T$-$Q$ equation) through the
Baxter vector on the spectral curve. In Sect. 3,
we consider the case when the spectral data lie on
rational curves
and perform the mathematical  derivation of the
answer. We present the complete solutions of  the 
Bethe polynomial equations
of all sectors for $L \leq 3$. In Sect. 4, we
discuss the "degeneracy" relation between the 
Bethe solutions and
the eigenspaces in the quantum space of the 
transfer
matrix for $L=3$; also its connection with the usual 
Bethe ansatz technique in literature, in
particular the result obtained in \cite{FK}. In
Sect. 5, we describe the algebraic geometry
properties of the high genus spectral curve
arisen from the Hofstadter Hamiltonian.

{\bf Notations. } The letters
$\ZZ,
\RZ, \CZ$ will denote  the ring of integers,
real, complex numbers respectively, $\NZ =
\ZZ_{>0}$, $\ZZ_N=
\ZZ/N\ZZ$. 
Throughout this report, $N$ will always denote an
odd positive  integer with $M= [\frac{N}{2}]$:
$N= 2M+1, M \geq 1$; $\omega$ is a primitive
$N$-th root of unity,  and $q:=
\omega^{\frac{1}{2}}$ with   
$q^N = 1$, i.e., 
$q = \omega^{M+1 }$. An element 
$v $ in the vector space 
$\CZ^N$ is represented by a sequence of
coordinates, $v_k ,  k
\in \ZZ$, with the $N$-periodic condition, $v_k =
v_{k+N}$, i.e., $v = (v_k)_{k
\in \ZZ_N}$.  
The standard basis of $\CZ^N$ will be denoted by 
$|k\rangle$, with the dual basis of $\CZ^{N*}$ by
$\langle k|$ for $k \in \ZZ_N$.
For a positive integers $n$, we denote
$\stackrel{n}{\otimes} \CZ^N$ the tensor
product of $n$-copies of the vector space $\CZ^N$.
We use the notation of $\rho$-shifted 
factorials: $
(a ; \rho )_n = 
(1-a)(1-a \rho) \cdots (1-a
\rho^{n-1}) $ for $ n \in \NZ$, and $(a ; \rho)_0
= 1$.

\section{Transfer Matrix and the Bethe Equation}
We consider the Weyl algebra generated by the
operators $Z, X$   satisfying the Weyl
commutation relation with the
$N$-th power identity, $
ZX= \omega XZ, Z^N = X^N = I$, and denote $Y :
=ZX$. In the canonical irreducible 
representation of the Weyl algebra,
the operators $Z, X, Y$ act on $\CZ^N$ with the
expressions:
$Z(v)_k = \omega^k v_k $, $
X(v)_k = v_{k-1} $, $Y(v)_k = \omega^k
v_{k-1}$. It is known that the following 
$L$-operator for an element $h=[a:b:c:d]$ of the projective 3-space 
$\PZ^3$ with operator-entries
acting on the quantum space
$\CZ^N$,    
$$
L_h (x) = \left( \begin{array}{cc}
       aY  & xbX  \\
        xcZ &d    
\end{array} \right) \ , \ \ x \in \CZ \ ,
$$
possesses the intertwining property of the Yang-Baxter
relation, 
\begin{eqnarray}
R(x/x') (L_h (x) \bigotimes_{aux}1) ( 1
\bigotimes_{aux} L_h (x')) = (1
\bigotimes_{aux} L_h (x'))(L_h (x)
\bigotimes_{aux} 1) R(x/x') \ , \label{eq:RLL}
\end{eqnarray}
where  
$R(x)$ is the matrix of a 2-tensor of the auxiliary 
space $\CZ^2$ with the following  numerical
expression,
$$
R(x) = \left( \begin{array}{cccc}
        x\omega-x^{-1}  & 0 & 0 & 0 \\
        0 &\omega(x-x^{-1}) & \omega-1 &  0 \\ 
        0 & \omega-1  &x-x^{-1} & 0 \\
     0 & 0 &0 & x\omega-x^{-1}     
\end{array} \right) \ .
$$
By performing the matrix product on 
auxiliary  spaces and the tensor product of quantum 
spaces,
one has the $L$-operator associated to an element 
$\vec{h}= (h_0,
\ldots, h_{L-1}) \in  (\PZ^3)^L $,
$
L_{\vec{h}}(x) =
\bigotimes_{j=0}^{L-1}L_{h_j}(x) $, 
which again satisfies the relation 
(\ref{eq:RLL}). The entries of $L_{\vec{h}}(x)$ are
operators of the quantum space 
$\stackrel{L}{\otimes} \CZ^N$, and its trace 
defines the commuting 
transfer matrices for $x \in \CZ$, $
T_{\vec{h}} (x) = {\rm tr}_{aux} ( L_{\vec{h}}(x))
$.
The transfer matrix $T_{\vec{h}} (x)$ can also be
computed by changing
$L_{h_j}$ to
$\widetilde{L}_{h_j}$ via a gauge transformation
:$
\widetilde{L}_{h_j} (x,
\xi_j, \xi_{j+1})= A_j L_{h_j}(x) A_{j+1}^{-1},
0 \leq j \leq L-1, $ with $
A_j = \left( \begin{array}{lc}
1 &  \xi_j-1\\
1 &  \xi_j 
\end{array}\right)$ and $ A_{L}: = A_0$. One has
$$
\widetilde{L}_{h_j} (x, \xi_j, \xi_{j+1})  = \left(
\begin{array}{ll} F_{h_j}(x,  \xi_j -1, \xi_{j+1}) &
-F_{h_j} (x, 
\xi_j -1, \xi_{j+1} -1) \\ F_{h_j}(x,  \xi_j ,
\xi_{j+1}) & - F_{h_j}(x, 
\xi_j ,
\xi_{j+1}-1 )
\end{array}\right) \ ,
$$
where $F_h(x,  \xi, \xi' ) :=
\xi'aY - xbX +  \xi'
\xi xc Z -\xi d$. Hence $
T_{\vec{h}} (x) = {\rm tr}_{aux}
(\widetilde{L}_{\vec{h}}(x,
\vec{\xi} ) ), \vec{\xi} := 
(\xi_0,  \ldots ,
\xi_{L-1} )$
where 
\begin{eqnarray*}
\widetilde{L}_{\vec{h}}(x,
\vec{\xi} ) : = 
\bigotimes_{j=0}^{L-1}
\widetilde{L}_{h_j}(x,
\xi_j, \xi_{j+1})   = 
\left( \begin{array}{cc}
   \widetilde{L}_{ \vec{h}; 1,1}(x, \vec{\xi}) &
\widetilde{L}_{
\vec{h};1,2}(x, \vec{\xi})  \\
     \widetilde{L}_{ \vec{h}; 2,1}(x, \vec{\xi}) &
\widetilde{L}_{ \vec{h} ; 2,2}(x, \vec{\xi}) 
\end{array} \right) , \ \ \ \ \xi_L := \xi_0 \ .
\end{eqnarray*}

We consider the variables $(x, \xi_0,
\ldots, \xi_{L-1})$ in the following spectral
curve, 
\begin{eqnarray}
{\cal C}_{\vec{h}} : \ \ \xi_j^N  =(-1)^N
\frac{\xi_{j+1}^Na_j^N - x^Nb_j^N}{\xi_{j+1}^N x^N
c_j^N - d_j^N } \ \ , \ \ \ \  j =0,
\ldots, L-1 , \label{eq:Cvh}
\end{eqnarray}
and denote $p_j = (x, \xi_j,
\xi_{j+1})$. Then the operator $F_{h_j}(x,  \xi_j,
\xi_j)$ has 1-dimensional null space in $\CZ^N$
generated by the vector 
$|p_j \rangle$ with the form:
$$
\langle 0| p_j \rangle = 1 \ , \ \ \ \
\frac{\langle m|p_j \rangle}{\langle
m-1|p_j\rangle} = 
\frac{\xi_{j+1}a_j
\omega^m  - xb_j }{
- \xi_j (  \xi_{j+1} x c_j \omega^m - d_j) }  \ .
$$
The Baxter vector 
$|p\rangle$ for $p \in {\cal C}_{\vec{h}}$ is 
defined by $
|p\rangle \ : = |p_0\rangle \otimes \ldots \otimes
|p_{L-1}\rangle 
\in \stackrel{L}{\otimes} \CZ^N $, which possesses
the following property:
\begin{eqnarray*}
\widetilde{L}_{\vec{h}; 1,1}(x, \vec{\xi})|p\rangle = 
|\tau_- p\rangle 
\Delta_-(p) , &
\widetilde{L}_{\vec{h}; 2,2}(x, \vec{\xi})|p\rangle = 
|\tau_+ p\rangle\Delta_+(p) \ , &
\widetilde{L}_{\vec{h}; 2,1}(x, \vec{\xi})|p\rangle = 0
\ , 
\end{eqnarray*}
where $\Delta_\pm$ are functions of 
${\cal
C}_{\vec{h}}$ defined by $\Delta_-(x, \vec{\xi}) 
=
\prod_{j=0}^{L-1}( d_j-x
\xi_{j+1} c_j )$, $\Delta_+(x, \vec{\xi}) = 
\prod_{j=0}^{L-1} \frac{\xi_j
(a_jd_j-x^2b_jc_j)}{\xi_{j+1}a_j -xb_j} $,
and $\tau_\pm$ are the automorphisms, $ \tau_\pm
(x, \vec{\xi})  = (q^{\pm 1} x, 
q^{-1} \vec{\xi}) $.
It follows the important relation of 
the transfer matrix on the Baxter vector over the curve 
${\cal C}_{\vec{h}}$, 
\bea(l)
T_{\vec{h}}(x) |p\rangle = |\tau_- p\rangle 
\Delta_-(p)  + |\tau_+ p\rangle\Delta_+(p) \ , \ \ 
{\rm for } \ \ p \in {\cal C}_{\vec{h}} \ .
\elea(T|p)
As $T_{\vec{h}}(x)$ are commuting operators for
$x \in \CZ$, 
a common eigenvector $\langle \varphi|$ is a constant 
vector of $\stackrel{L}{\otimes} \CZ^N$ 
with an eigenvalue
$\Lambda(x) \in \CZ[x]$. Define the function
$Q(p)= \langle \varphi|p\rangle$ of ${\cal
C}_{\vec{h}}$, then it 
 satisfies the
following Bethe equation, 
\bea(l)
\Lambda(x) Q(p)  = Q(\tau_-(p)) \Delta_-(p) 
+  Q(\tau_+(p)) \Delta_+(p) \ , \ \ {\rm for} \ p 
\in {\cal C}_{\vec{h}} \ .
\elea(Bethe)
By the definition of $T_{\vec{h}}(x), 
\Lambda(x)$, one can easily see that
$T_{\vec{h}}(x)$ is an operator-coefficient even
$x$-polynomial of degree $2[\frac{L}{2}]$
with the constant term $
T_0 =  \prod_{j=0}^{L-1} a_j 
\stackrel{L}{\bigotimes} Y + 
\prod_{j=0}^{L-1} d_j $. Hence
the polynomial $\Lambda (x)$ in {\rm
(\req(Bethe))} is an  even function of
degree
$\leq 2[\frac{L}{2}]$ with $\Lambda (0) = 
q^l \prod_{j=0}^{L-1} a_j + \prod_{j=0}^{L-1} d_j  $ 
for some $l \in \ZZ_N$. For  $L=3$,  we have 
$T_{\vec{h}}(x)= T_0 + x^2 T_2 $ where
\bea(ll)
  T_2 =& b_0c_1a_2 X\otimes Z \otimes Y +
a_0b_1c_2 Y \otimes  X \otimes Z+ 
 c_0a_1b_2 Z \otimes Y
\otimes X \\
&+ \ c_0b_1d_2 Z\otimes X \otimes I +
d_0c_1b_2 I \otimes  Z \otimes X+ 
 b_0d_1c_2 X \otimes I
\otimes Z \ .
\elea(T02)
The above $T_2$  can be put into the form
of the
 Hofstadter type Hamiltonian (\req(HFK)) \cite{FK} \cite{LR}.

In the equation (\req(Bethe)), $Q(p)$ is a
 rational function  of ${\cal C}_{\vec{h}}$
with zeros and poles. 
Hence the understanding  of the Bethe solutions of
(\req(Bethe)) relies heavily on the function theory of
${\cal C}_{\vec{h}}$, and the algebraic geometry of the
curve inevitably plays a key role on the
complexity of the problem.

\section{The Rational Degenerated Bethe Equation}
In this section, we consider the case when the
spectral curve 
${\cal C}_{\vec{h}}$ degenerates into an union of 
rational  curves under the conditions:$
a_j = q^{-1} d_j ,  \ b_j = q^{-1} c_j $ for $
j=0, \ldots, L-1$. 
By replacing $c_j,
d_j$  by $\frac{c_j}{d_j}, 1$, we 
assume $ d_j=1$ for all $j$ with  the parameter
$c_j$s to be generic.  
In this case,  ${\cal C}_{\vec{h}}$ is the 
union of disjoint copies of the $x$-(complex) line
, containing the following
$\tau_\pm$-invariant subset of 
${\cal C}_{\vec{h}}$ which will
be sufficient for the discussion of Bethe
equation,
$$
{\cal C} := \{ (x, \xi_0, \ldots, \xi_{L-1})  |  
 \xi_0 = \cdots = \xi_{L-1} =  q^l ,  l \in
\ZZ_N \} .
$$ 
We shall make the identification $
{\cal C} = \PZ^1 \times \ZZ_N $ via $ 
(x,  q^l, \ldots,  q^l)
\leftrightarrow  (x, l) $. The automorphisms $\tau_\pm$ on ${\cal C}$  
  become $
\tau_\pm (x, l)= (q^{\pm 1} x, l-1) $, 
by which the action (\req(T|p)) of $T(x) (:=
T_{\vec{h}}(x))$ on the Baxter vector $|x,
l\rangle$ now takes the form,
\bea(l)
T (x) |x, l\rangle = |q^{-1}x, l-1\rangle 
\Delta_-(x, l)  + |qx, l-1\rangle\Delta_+(x, l) \  , 
\elea(T|xm)
where $\Delta_\pm$ are the rational
functions of $x$:$
\Delta_-(x, l)  = 
\prod_{j=0}^{L-1}( 1-x
   c_j q^l) $, $
\Delta_+(x, l) = 
\prod_{j=0}^{L-1} \frac{
1-x^2c^2_j}{1 -x c_j q^{-l} }$.
Furthermore,  one can express the Baxter vector 
$|x, l\rangle$ 
over the curve ${\cal C}$ in the 
component-form : $
\langle {\bf k}|x, l \rangle  = q^{|{\bf k}|^2}
\prod_{j=0}^{L-1} 
\frac{ (x c_jq^{-l-2}; \omega^{-1})_{k_j} }{ (x 
c_j q^{l+2}; \omega )_{k_j}}$.
Here the bold letter ${\bf k}$ denotes a
multi-index vector
${\bf k}= (k_0, \ldots, k_{L-1})$ for 
$k_j \in \ZZ_N$ with the square-length of ${\bf
k}$ defined by 
$|{\bf k}|^2:= \sum_{j=0}^{L-1} k_j^2$.  
Each ratio-term  in the above right
hand side is given by a non-negative 
representative for each element in $\ZZ_N$ 
appeared in the formula. We have the following 
result on the Bethe equation and its  connection
with the transfer matrix 
$T(x)$:
\begin{theorem}  
Denote $f^e, f^o$ the functions on 
${\cal C}$, $
f^e(x, 2n) =  \prod_{j=0}^{L-1}
\frac{(xc_j ; \omega^{-1} )_{n+1} }{
(xc_j; \omega )_{n+1}} $, and $
f^o(x, 2n+1) =  \prod_{j=0}^{L-1}
\frac{(xc_jq^{-1}; \omega^{-1} )_{n+1}}{
(xc_jq ; \omega)_{n+1}} $. 
For $x \in \PZ^1$ , $ l \in \ZZ_N$,
we define the following vectors in
$\stackrel{L}{\otimes}\CZ^N$, 
\begin{eqnarray*}
|x\rangle_l^e = \sum_{n=0}^{N-1}  |x, 2n\rangle  
f^e (x, 2n)
\omega^{ln} 
 \ , &
|x\rangle_l^o = \sum_{n=0}^{N-1} 
 |x, 2n+1\rangle f^o (x, 2n+1)  \omega^{ln}  \ , 
\\
|x \rangle_l^+ =  |x\rangle_l^e q^{-l} u(qx) +  
|x \rangle_l^o u(x) ;& \ \ {\rm where} \ 
u(x): = \prod_{j=0}^{L-1} (1-x^Nc^{
N}_j)(xc_jq ; q^2)_M \ .
\end{eqnarray*} 
Then 

(i) $|x \rangle_l^e u(qx) = 
|x\rangle_l^o q^lu(x)$, or equivalently, 
$
|x\rangle_l^+ = 2 q^{-l} |x\rangle_l^e u(qx)  = 2
|x\rangle_l^ou(x) $. 

(ii) The $T(x)$-transform on $| x\rangle^+_l$ is
given by
$$
q^{-l} T(x)|x\rangle^+_l
=|q^{-1}x\rangle^+_l  \Delta_-(x, -1) +
|qx\rangle^+_l  \Delta_+(x, 0) \ , \ \ \ l \in \ZZ_N \ .
$$

(iii) For a common eigenvector $\langle \varphi|$
of $T(x)$ with the  eigenvalue
$\Lambda (x)$, the function
$Q_l^+(x)  (:=\langle \varphi | x\rangle^+_l)$
and $\Lambda (x)$ are  polynomials with the
properties: $
{\rm deg.}Q^+_l (x)
\leq  (3 M +1)L$ , $
 {\rm deg.}
\Lambda (x) \leq 2[\frac{L}{2}]$, $
\Lambda (x)= \Lambda(-x)$, $ 
\Lambda (0)= 
q^{2l}+1  $, 
and the following Bethe equation
holds:
\begin{eqnarray}
 q^{-l} \Lambda (x) Q_l^+ (x) = 
\prod_{j=0}^{L-1}(1-xc_jq^{-1})
Q_l^+ ( xq^{-1}) + \prod_{j=0}^{L-1} 
(1+x c_j) Q_l^+ (xq) \ .
\label{QmEq}
\end{eqnarray}
Furthermore for $0 \leq m \leq M$, 
$Q_m^+(x), Q_{N-m}^+(x)$ are elements in 
$ x^m \prod_{j=0}^{L-1} (1-x^Nc^{ N}_j)
\CZ[x]$. 
\end{theorem}
By (iii) of the above theorem,
the  equation (\ref{QmEq}) for the sector $m,
N-m$ can be combined into a single one. 
For the rest of this report the letter $m$ will
always denote an integer between 
$0$ and $M$: $
0 \leq m \leq M$. 
By introducing the 
polynomials
$\Lambda_m(x), Q ( x)$ via the relation,
$$
(\Lambda_m(x) , \ x^m
\prod_{j=0}^{L-1} (1-x^Nc^{ N}_j) Q ( x) ) = 
(q^{-m}\Lambda(x), \ Q_m^+(x) ) , \ \ (q^m
\Lambda (x) , \ Q_{N-m}^+(x) )  \ ,
$$
the equation (\ref{QmEq}) for $l=m, N-m$  
becomes the  following 
polynomial equation of $Q (x), \Lambda_m(x)$: 
\be
 \Lambda_m (x) 
Q (x) = q^{-m} 
\prod_{j=0}^{L-1}(1-xc_jq^{-1}) 
Q (xq^{-1})+ q^m \prod_{j=0}^{L-1} (1+x c_j) 
 Q (xq) \ ,
\ele(rBeq)
with $
{\rm deg.}Q (x)
\leq ML-m$, $ 
 {\rm deg.}
\Lambda_m (x) \leq 2[\frac{L}{2}]$, $ 
\Lambda_m (x)= \Lambda_m (-x),  \Lambda_m
(0)=  q^m + q^{-m}$.
The general mathematical problem will be  to
determine  the solution space of the Bethe
equation (\req(rBeq)) for a given positive integer
$L$.

For $L=1, 2$, we have the following result.
\begin{theorem} (I)
For $L=1$, we have $\Lambda_m (x)=  q^m+q^{-m}$
and the solutions  
$Q_m(x)$ of ${\rm (\req(rBeq))}$ form an
one-dimensional vector space  generated by the following 
polynomial of degree $M-m$,
$$
B_m(x) = 1+ \sum_{j = 1}^{M-m} (\prod_{i=1}^j 
\frac{q^{m+i-1} - q^{-m-i} }{q^m+q^{-m}-q^{-m-i}-q^{m+i}})
(xc_0)^j \ .
$$

(II) For $L=2$, 
the equation ${\rm (\req(rBeq))}$ has a
non-trivial solution $Q_m(x)$ if and only if ${\rm deg.}
Q_m(x) = M -m +m'$ for $0 \leq m' \leq M$. For each such
$m'$, the eigenvalue $\Lambda_m(x)$ in ${\rm
(\req(rBeq))}$ is equal to $
\Lambda_{m,m'}(x) :=  q^{\frac{1}{2}} (q^{m'-1} + 
q^{-m'-2})x^2 c_0c_1  +q^m+q^{-m}$, and the
corresponding solutions of $Q_m(x)$  form an
one-dimensional space generated by a
 polynomial $B_{m,m'}(x)$ of degree $M-m+m'$ with 
$B_{m,m'}(0) = 1$.
\end{theorem}

For $L=3$, this is the case related to
the Hamiltonian (\req(HFK)). We consider  the
$N
\times N$ matrix,
\bea(l)
A=  \left( \begin{array}{ccccccc}
\delta_{N-1}' & u_{N-1}' &0&\cdots  &  &0 &0
\\ v_{N-2}'& \delta_{N-2}' 
&u_{N-2}'&0&\ddots  &  &\vdots \\
w_{N-3}'&v_{N-3}'& \delta_{N-3}' 
&u_{N-3}'&\ddots &  & 0  \\
0& \ddots&\ddots &\ddots  & \ddots & &\vdots \\
\vdots& \ddots & \ddots & \ddots & \ddots &\ddots  &0\\
\vdots& \ddots &0&w_1' & v_1'  & 
\delta_1' &u_1'\\
0&  \cdots&  & 0 & w_0' & v_0' &
\delta_0'
\end{array} 
\right) 
\elea(AL3)
with the entries defined by
$
w_k'= q^{k+\frac{3}{2}}
+q^{-k-\frac{3}{2}}-q^m-q^{-m}$, $  
v_k' =
(q^{k+\frac{1}{2}}
-q^{-k-\frac{3}{2}})(c_0+c_1+c_2)$, $
\delta_k' = (q^{k-\frac{1}{2}}
+q^{-k-\frac{3}{2}}) (c_0c_1+c_1c_2+c_2c_0) $, $
u_k'=
(q^{k-\frac{3}{2}}-q^{-k-\frac{3}{2}})c_0c_1c_2
$.  Then one can derive the following result.
\begin{theorem}\label{thm:L3sol}
For $L=3$, the condition of the eigenvalue  
$\Lambda_m(x)= \lambda_m x^2 +
q^m + q^{-m} $, $0 \leq m \leq M$, with a
non-trivial solution
$Q_m(x)$ in the equation
${\rm (\req(rBeq))}$ is determined by the
solution of
${\rm det}(A-\lambda_m) =0$, where $A$ is the matrix
defined by
$(\req(AL3))$. For
each such 
$\Lambda_m(x)$, there exists an unique (up to constants)
non-trivial polynomial solution 
$Q_m(x)$ of 
${\rm (\req(rBeq))}$ with the degree $Q_m(x)$
equal to $3M-m$ and $Q_m(0) \neq 0$.
\end{theorem}

\section{The Degeneracy and  Bethe Ansatz
Relation of Roots of Bethe Polynomial}  
We first discuss the
degeneracy relation of eigenspaces of the transform
matrix
$T(x)$ in
$\stackrel{L}{\otimes}\CZ^{N*}$ with respect to the
Bethe solutions obtained in the previous section.
As before, we denote $\Lambda (x)$ the
eigenvalues of $T(x)$, whose constant term is
given by $
T_0  = D + 1$, where $ D := q^{-L}
\stackrel{L}{\otimes} Y $; hence
$\Lambda(0)= q^l+1$. For $l \in \ZZ_N$, we denote 
$\EZ_L^l$  the $N^{L-1}$-dimensional eigensubspace
of
$\stackrel{L}{\otimes}\CZ^{N*}$ of 
the operator $D$ with the eigenvalue $ q^l$.
For $0 \leq m \leq M$, the equation
(\req(rBeq)) describes the relation of   
$\Lambda(x)$ and its eigenfunctions with 
$\Lambda(0)= q^{2m}+1$ or $ q^{2(N-m)}+1$.
We now consider the case for $L=3$, where 
 $T_2$ in (\req(T02)) is now expressed by
$$
\begin{array}{ll}
T_2=& q^{-2}( c_0c_1  X\otimes Z \otimes Y +
 c_1c_2 Y \otimes  X \otimes Z+ c_0c_2 Z
\otimes Y
\otimes X ) \\
&+
 q^{-1} ( c_0c_1 Z\otimes X
\otimes I + c_1c_2 I \otimes  Z \otimes X + 
c_0c_2 X
\otimes I
\otimes Z ) .
\end{array}
$$
We have $
qD = (Z\otimes X \otimes I)( X
\otimes I \otimes Z)( I
\otimes Z \otimes X)$.
We shall denote ${\cal O}_3$ the operator
algebra generated by the tensors of 
$X, Y, Z, I$
appeared in the above expression of $T_2$. Then 
${\cal O}_3$ 
commutes with 
$D$, hence one obtains a ${\cal
O}_3$-representation on
$\EZ_3^l$ for each $l$. 
With the identification, $
U  = D^{-1/2}  Z\otimes X \otimes I $, $ V
=  D^{-1/2}  X
\otimes I \otimes Z  $, 
${\cal O}_3$ is generated by $U, V$ which satisfy the
Weyl relation
$UV= \omega VU$ and  the $N$-th power identity.
Hence 
${\cal O}_3$ is the Heisenberg algebra and contains 
$D$ as a central element.  Then  
$qD^{\frac{-1}{2}}T_2$ has the following
expression,
\bea(l)
c_0c_1 (   U  +
U^{-1} )
 +  c_0c_2 ( V +
V^{-1} ) +  c_1c_2 
( q 
D^{5/2}  UV + q^{-1}  D^{-5/2}  V^{-1}U^{-1} ) \ .
\elea(GSHof)
The above Hamiltonian is the same as $H_{FK}$
(\req(HFK)) with
$W= q^{-1}  D^{-5/2}  V^{-1}U^{-1} $,
$\alpha=\beta=
\gamma = 1$. Our conclusion on the sector $m=M$ is
equivalent to that in
\cite{FK} as it  becomes clearer later on. 
There is an unique (up to  equivalence) 
non-trivial irreducible representation of ${\cal O}_3$,
denoted by
$\CZ^N_{\rho}$, which is
of dimension $N$. For each $l$, 
$\EZ_3^l$ is equivalent to $N$-copies of
$\CZ^N_{\rho}$ as ${\cal
O}_3$-modules:  
$\EZ_3^l \simeq N
\CZ^N_{\rho}$. For $0 \leq m \leq M$, we consider the
space
$\EZ_3^l$ with 
$q^l= q^{\pm 2m}$. The evaluation of $\EZ_3^l$
on $|x\rangle_{\pm m}^+$ gives rise to a $N$-dimensional
kernel in
$\EZ_3^l$. By Theorem
\ref{thm:L3sol}, there are $N$ 
polynomials $Q_m(x)$ of degree $3M-m$  as  solutions
of 
${\rm (\req(rBeq))}$  with the corresponding
$N$ distinct eigenvalues $\Lambda_m(x)$. The $N$-dimensional
vector space spanned by those 
$Q_m(x)$s becomes a realization of the
irreducible  representation
$\CZ^N_{\rho}$ for the Heisenberg algebra
${\cal O}_3$.

Now we discuss the relation between the Bethe
equation (\req(rBeq)) and the usual  Bethe
ansatz formulation in literature.  For 
$0\leq m \leq M$, a solution $Q_m(x)$ in
$(\req(rBeq))$ always have the property $Q_m(0) \neq
0$ by Theorem \ref{thm:L3sol}, hence one has the form 
$
\alpha_{3M-m}^{-1} Q_m (x) = \prod_{l=1}^{3M-m} ( x -
\frac{1}{z_l} ) $ with $ 
\  z_l \in \CZ^* $. 
By setting $x= z_l^{-1}$ in 
${\rm (\req(rBeq))}$ , we obtain the
following relation among
$z_l$s, which is called the Bethe
ansatz relation,
$$
 q^{m+\frac{3}{2}}  \prod_{j=0}^2 \frac{z_l +
 c_j}{qz_l - c_j} =   \prod_{n=1, n \neq l}^{3M-m}
\frac{ qz_l - z_n }{ 
z_l -  q z_n  } ,  \ 1 \leq l \leq 3M-m \ .
$$
For the sector $m=M$, 
the comparison of the $x^2$-coefficients of ${\rm
(\req(rBeq))}$ yields the expression of
eigenvalue,
$$
 \lambda_M 
= ( q^{\frac{-1}{2}}+ q^{\frac{-3}{2}} )s_2 
+  (q^{\frac{1}{2}} - q^{\frac{-3}{2}}) s_1
\sum_{n=1}^{2M} z_n
 + 
(q^{\frac{3}{2}} + q^{\frac{-3}{2}}  - q^{\frac{1}{2}}
-q^{\frac{-1}{2}} ) 
\sum_{l<n} z_lz_n  \ . 
$$
With the substitution, $
\mu = q^{\frac{1}{2}}c_0^{-1}, 
\nu = q^{\frac{1}{2}}c_1^{-1},  
\rho= q^{\frac{1}{2}}c_2^{-1} $, the
above expression coincides with  
(5.27) in  \cite{FK}. Note that the
Bethe ansatz relation can be shown to
be equivalent to the Bethe equation
$(\req(rBeq))$ for the sector $M$. However,
the parallel statement is no longer true for
other sectors 
$m \neq M$, i.e., it does exist some non-physical
 Bethe ansatz solutions in the above form, while
not corresponding to any polynomial solution
of Bethe equation (\req(rBeq)). Some  
example can be found in the 
$(M-1)$-sector.

\section{High Genus Curves for the Hofstadter
Model} We are now going back to the general
situation in Sect. 2. Note that the values
$\xi_j^N$s of the curve
${\cal C}_{\vec{h}}$ in (\ref{eq:Cvh}) are determined
by $\xi_0^N$ and $x^N$, denoted by $
y = x^N $, $ \eta =\xi_0^N$.
The variables $(y, \eta)$ defines the curve which is a double cover of 
$y$-line,
$$
{\cal B}_{\vec{h}} : \ \ C_{\vec{h}}(y) \eta^2 +
(A_{\vec{h}}(y)-  D_{\vec{h}}(y))\eta 
- B_{\vec{h}}(y) = 0 
$$
where the functions $A_{\vec{h}}, B_{\vec{h}}, 
C_{\vec{h}}, D_{\vec{h}}$  are the following
matrix elements,  
$$
\left( \begin{array}{cc}
-A_{\vec{h}}(y)  &  B_{\vec{h}}(y)  \\
C_{\vec{h}}(y)  & - D_{\vec{h}}(y)    
\end{array} \right) :
= \prod_{j=0}^{L-1}  \left( \begin{array}{cc}
- a^N_j  &  y b^N_j  \\
 y c^N_j  &-d^N_j    
\end{array} \right) \ .
$$
Now we consider only the case: $
L = 3$, $ a_0=d_0=0 , b_0=c_0 = 1$, 
with generic $h_1, h_2$.  The expression of
$T(x)$ is given by
$$
 T( x)  = x^2 (c_1a_2 X\otimes Z \otimes Y + 
 a_1b_2 Z \otimes Y
\otimes X  + b_1d_2 Z\otimes X \otimes I + 
 d_1c_2 X \otimes I
\otimes Z ) ,
$$
equivalently, $x^{-2} D^{\frac{-1}{2}}  T( x)$ is
equal to the  Hofstadter Hamiltonian
$(\req(HFK))_{\rho=0}$ with
$U  = D^{-1/2}  Z\otimes X \otimes I $, $ V
=  D^{-1/2}  X
\otimes I \otimes Z  $ and 
 $\mu, \nu,
\alpha,
\beta$ related to 
$h_1, h_2$  by $
\mu^2 = qb_1c_1a_2d_2$, $\alpha^2 =
q^{-1}b_1c_1^{-1}a_2^{-1}d_2$, $ \nu^2 = q
a_1d_1b_2c_2 $, $ 
\beta^2= q^{-1}a_1^{-1}d_1b_2^{-1} c_2$. 
By factoring out the $y$-component of ${\cal
B}_{\vec{h}}$, the main
irreducible component of
${\cal B}_{\vec{h}}$ is the curve,
$$
{\cal B}: \ (y^2b_1^Nc_2^N + a^N_1a_2^N ) 
\eta^2 + ( a_1^N b^N_2+ b_1^Nd_2^N 
- c^N_1a_2^N- d_1^N
c_2^N) y \eta  -
(y^2c_1^Nb_2^N+d_1^Nd^N_2 ) = 0 \ ,
$$
which is an elliptic curve as a double-cover of
the $y$-line. For the curve ${\cal
C}_{\vec{h}}$, 
the variables $\xi_0$ and $ \xi_1$ are related by
$\xi_0^N =
\xi_1^{-N}$, which implies that ${\cal
C}_{\vec{h}}$ can be identified with ${\cal W}
\times \ZZ_N $ where ${\cal W}$ is a genus 
$6N^3-6N^2+1$ curve
 with the following
equation in the variable
$p=(x,\xi_0,\xi_2)$, 
$$
{\cal W} : \ \ \xi_0^{-N} = \frac{-\xi_2^Na_1^N+
x^Nb_1^N}{x^N\xi_2^Nc_1^N-d_1^N} \ , \ \ \
\xi_2^N= \frac{-\xi_0^Na_2^N+
x^Nb_2^N}{x^N\xi_0^Nc_2^N-d_2^N} \ .
$$
By averaging the Baxter vectors $|p, s \rangle$
of ${\cal
C}_{\vec{h}} $ over an element
$p$ of ${\cal W}$, $
|p \rangle:= \frac{1}{N} \sum_{s=0}^{N-1} |p, s \rangle
q^{s^2}$, which defines the 
Baxter vector on 
${\cal W}$. 
Furthermore, the transfer matrix can  be
descended to one on ${\cal W}$ with the following
relation,
$$
x^{-2}T(x)|p\rangle = |\tau_-(p)\rangle
\widetilde{\Delta}_-(p) +  |\tau_+(p)\rangle 
\widetilde{\Delta}_+(p) \ ,
$$
where $\widetilde{\Delta}_{\pm}$ are the
functions on
${\cal W}$: $
\widetilde{\Delta}_-(x, \xi_0, \xi_2 ) =
 \frac{( x
\xi_2 c_1 -d_1 ) ( x
\xi_0 c_2 - d_2 )}{- x \xi_0}$, $
\widetilde{\Delta}_+(x, \xi_0, \xi_2 ) = 
\frac{ \xi_2
(a_1d_1-x^2b_1c_1)(a_2d_2-x^2b_2c_2)}{x (\xi_2a_1
-xb_1)(\xi_0a_2-xb_2)}$. For an eigenvector $\langle \varphi|
\in \stackrel{3}{\otimes} \CZ^{N*}$ of 
$x^{-2}T(x)$, 
the eigenvalue is a scalar
$\lambda \in \CZ$, and the function 
$Q(p) : = \langle  \varphi |
p \rangle  $ 
 of ${\cal W}$ satisfies the  Bethe
equation: $
\lambda Q(p) = Q(\tau_-(p)) \widetilde{\Delta}_-(p) + 
Q(\tau_+(p)) \widetilde{ \Delta}_+(p) $, 
where $\tau^\pm$ are the transformations of
${\cal W}$ with the same expression as before,
but only in the coordinates $(x, \xi_0, \xi_2)$.
Consider the
$D$-eigenspace decomposition
of 
$\stackrel{3}{\otimes} \CZ^{N*} = \bigoplus_{l\in
\ZZ_N} \EZ_3^l$.
The evaluation of the Baxter vector over ${\cal W}$ gives
rise  to the following linear transformation,
$
\varepsilon_l : \EZ_3^l \longrightarrow \{ {\rm 
rational \ functions \ of \ } {\cal W} \}$
with $\varepsilon_l(v)(p) := \langle v|p\rangle$, 
for $ l \in \ZZ_N$. 
One has the following result.
\begin{theorem}
For $l \in \ZZ_N$, the
linear map $\varepsilon_l$ is injective, hence it induces
an identification of $\EZ_3^l$ with a
$N^2$-dimensional functional space of
${\cal W}$.  
\end{theorem}
By the discussion in Sect. 4, as the Heisenberg
algebra
${\cal O}_3$ representations , $\EZ_3^l$ is
equivalent to
$N$ copies of the standard one. Hence it induces an
${\cal O}_3$-module structure on the function space
$\varepsilon_l(\EZ_3^l)$, induced by the one of
$\EZ_3^l$ by above theorem. The
mathematical structure of the functional space 
$\varepsilon_l(\EZ_3^l)$ in terms of the divisor theory
of Riemann surfaces in corporation with the
interpretation of Heisenberg algebra representation  
remains an  algebraic geometry problem for further study.

\section*{Acknowledgements}
S.S. Roan is grateful to  
M. Jimbo, J.  Kellendonk, B. M. McCoy , R. Seiler
for fruitful discussions. This work was reported
 in the workshop "Quantum
Algebra and Integrabilty" CRM Montreal, Canada,
April 2000, and was a subject of an
Invited Lecture at APCTP-Nankai Joint Symposium
on "Lattice Statistics and Mathematical Physics",
Tianjin, China, October 2001, to which he
would like to thank for their invitation and
hospitality.


\begin{thebibliography}{0}
\bibitem{AMP} G. Albertini, B. M. McCoy, and 
J. H. H. Perk, in {\it Adv. Stud. Pure Math.,
vol. 19}, ( Kinokuniya Academic 1989) p.1.

\bibitem{A} M.Ya. Azbel, {\it Sov.
Phys. JETP} {\bf 19}, 634 (1964).



\bibitem{BBP} R.J. Baxter, V.V. Bazhanov and J.H.H.
Perk,  {\it Int. J. Mod. Phys.} {\bf B4}, 803
(1990).


\bibitem{BazS} V.V. Bazhanov and Yu.G. Stroganov, 
{\it J.
Stat. Phys.} {\bf 59}, 799 (1990).

\bibitem{C} W.G. Chambers, {\it Phys. Rev.} {\bf
A140}, 135 (1965). 



\bibitem{FK} L. D. Faddeev and R. M. Kashaev, {\it
Comm. Math. Phys.} {\bf 155}, 181 (1995),
hep-th/9312133.


\bibitem{Har} P.G. Harper, {\it Proc.
Phys. Soc. London} {\bf A68}, 874 (1955).



\bibitem{H} D. R. Hofstadter, {\it Phys. Rev.} 
{\bf B14}, 2239 (1976).



\bibitem{L} D. Langbein, {\it Phys. Rev.} {\bf 180
}, 633 (1969).

\bibitem{LR} S. S. Lin and S. S. Roan,
''Algebraic geometry approach  to the Bethe
equation for Hofstadter type models'', (to appear
in J. Phys. A: Math. Gen. ), cond-mat/9912473.



\bibitem{P} R. Peierls, {\it Z. Phys.} {\bf 80},
763 (1933).




\bibitem{WZ} P. B. Wiegmann and A. V. Zabrodin,
{\it Nucl. Phys.} {\bf B422}, 495
(1994); {\it Nucl. Phys.} {\bf B451}, 699 (1995),
cond-mat/9501129.


\bibitem{Z} J. Zak, {\it Phys.
Rev.} {\bf A134}, 1602 (1964); {\it Phys. Rev.}
{\bf  A134}, 1607 (1964). 

\end{thebibliography}
\end{document}